\documentclass[conference]{IEEEtran}
\IEEEoverridecommandlockouts

\IEEEoverridecommandlockouts
\usepackage{graphicx}
\usepackage{algorithmic}
\usepackage{algorithm}
\usepackage[T1]{fontenc}
\usepackage[cmex10]{amsmath}
\usepackage{url}
\usepackage{multirow}

\title{Spoken Language Corpora Augmentation with Domain-Specific Voice-Cloned Speech\thanks{This research was partially funded by the \emph{CAIMAC: Conversational AI Multilingual Augmentation and Compression} project, a cooperation between Adam Mickiewicz University and Samsung Electronics Poland.}}
\author{
\IEEEauthorblockN{Mateusz Czy\.{z}nikiewicz, \L{}ukasz Bondaruk,\\ Jakub Kubiak, Adam Wi\k{a}cek, \L{}ukasz Deg\'{o}rski}
\IEEEauthorblockA{
Samsung R\&D Institute Poland\\
Plac Europejski 1\\ 00-844 Warszawa, Poland\\
Email: \{m.czyznikiew,l.bondaruk,j.kubiak3,\\a.wiacek2,l.degorski\}@samsung.com}
\and
\IEEEauthorblockN{Marek Kubis, Pawe\l{} Sk\'{o}rzewski}
\IEEEauthorblockA{0000-0002-2016-2598\\ 0000-0002-5056-2808\\
Adam Mickiewicz University, Poland\\
Faculty of Mathematics and Computer Science\\
ul. Uniwersytetu Poznanskiego 4\\
61-614 Poznan, Poland\\
Email: \{mkubis, pawel.skorzewski\}@amu.edu.pl}
}

\begin{document}
\maketitle

\begin{abstract}
In this paper we study the impact of augmenting spoken language corpora
with domain-specific synthetic samples
for the purpose of training a speech recognition system.
Using both a conventional neural TTS system and a zero-shot one with voice cloning ability
we generate speech corpora that vary in the number of voices.
We compare speech recognition models trained with addition of different
amounts of synthetic data generated using these two methods with a baseline model
trained solely on voice recordings.
We show that while the quality of voice-cloned
dataset is lower, its increased multivoiceity makes it much more effective than the one
with only a few voices synthesized with the use of a conventional neural TTS system.
Furthermore, our experiments indicate that using low variability synthetic speech
quickly leads to saturation in the quality of the ASR whereas high variability speech provides
improvement even when increasing total amount of data used for training by $30\%$.
\end{abstract}

\section{Introduction}
\IEEEPARstart{W}{ith the} development of better TTS systems in recent years, there has been
an increasing number of research papers on using synthesized data for ASR training
\cite{fazel2021synthasr, 9688218, rossenbach21comparing}.
One could argue that, if synthesized samples covered a more diverse set of voice
characteristics, even with decrease in speech quality, the data could be used more
effectively for training ASR.
Conventional neural TTS systems \cite{tan2021survey}, like Tacotron2 \cite{tacotron2}
or FastSpeech \cite{ren2019fastspeech}, require large amount of high-quality
paired text and speech data, which is not available for most languages,
especially for multiple voices. Because of that, we cannot use them to produce output with
more than a few to a dozen of voices, even for otherwise high-resource languages like German \cite{tan2021survey}.
Recent advancements in speech synthesis brought zero-shot models that use
neural codec encoding instead of mel-spectogram speech representation
\cite{wang2023neural, zhang2023vallex, shen2023naturalspeech}.
Thanks to their zero-shot voice cloning ability, they are able to generate high quality
audio with any person's voice, having just a few seconds recording of it. This allows for
generating synthetic corpora with hundreds of voices.

Our work examines the usefulness of having a synthetic corpora with a diverse set of voices.
For comparison, we employ a zero-shot TTS and a conventional neural TTS to produce a domain-specific
synthetic dataset with high and low number of speakers, respectively. We select a virtual assistant (VA)
domain as our experiment target. Then, we examine the usefulness of both synthetic datasets in improving
the ASR model's performance. We show that the high voice diversity of generated data
makes it much more effective.
Furthermore, our results indicate that
the potential for using synthesized data to improve the ASR
performance is limited by variability of the speech produced by a conventional neural TTS system.

\begin{figure*}[ht]
  \begin{center}
  \includegraphics[scale=0.65]{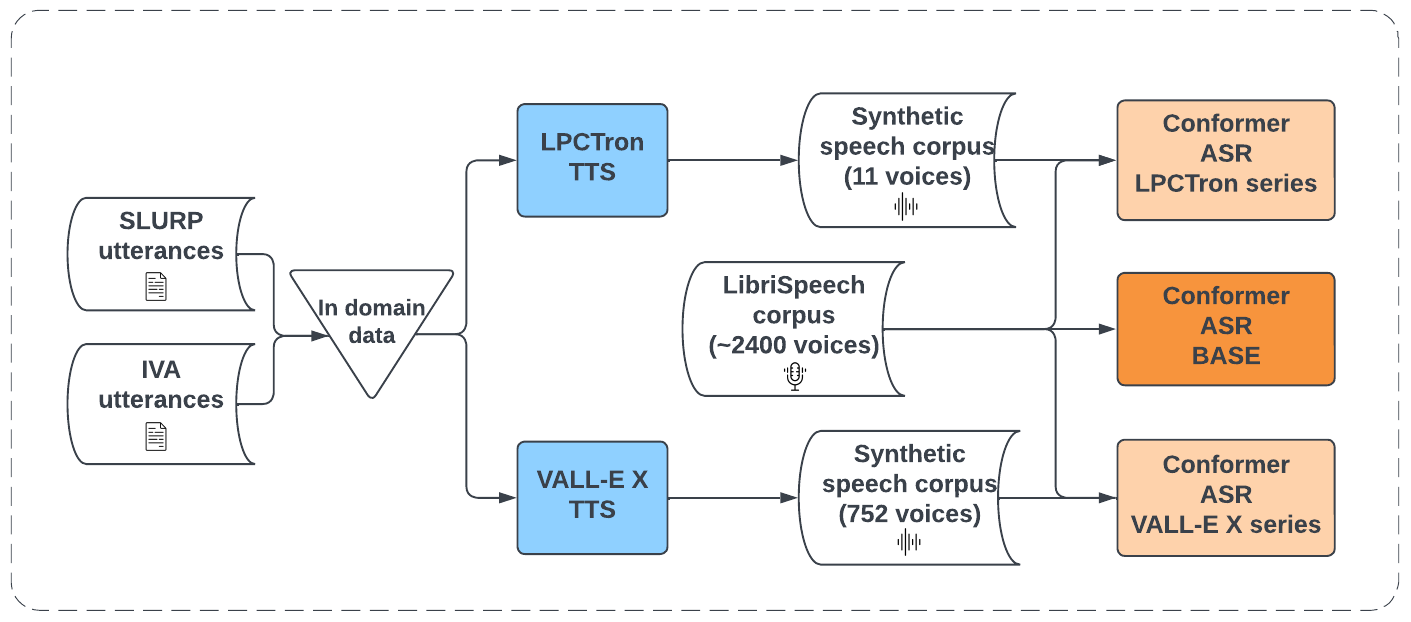}
  \caption{Experimental workflow.}
  \label{fig:workflow}
  \end{center}
\end{figure*}

\section{Related Work}

Prior work has shown that using text-to-speech data can improve ASR performance.
Rossenbach et al. \cite{rossenbach21comparing} examined the impact of synthetic data for various ASR architectures.
They showed that using TTS data pre-processing techniques can increase the robustness of ASR training.
They reported $38\%$ relative improvement after adding synthetic data to the attention encoder-decoder ASR system.

The addition of synthetic data can play an important role in a low-resource setting.
Bartelds et al. \cite{bartelds2023making} showed that adding synthetic data to the ASR training on
such languages like Besemah and Nasal reduced relative WER up to $25.5\%$.

In some situations, all that is needed to build an ASR is a text corpus.
Rossenbach et al. \cite{rossenbach19generating} demonstrated this strategy.
They achieved relative improvement of up to $33\%$ in WER over the baseline with
data augmentation in a low-resource setting.

Another use for synthetic data can be to improve the recognition of out-of-vocabulary (OOV)
words \cite{zheng2021using}. OOV is a prevalent issue encountered by
real-world virtual assistants that must adapt to the ever-evolving environment.
Augmentation using TTS-generated data for these specific OOV words can
positively affect the robustness of the ASR model without significant degradation
on the general dataset.

Kubis et al. \cite{kubis23back} use synthesized data
to study the impact of speech
recognition errors on the performance of natural language understanding models.
In \cite{kubis23caiccaic} text-to-speech models are used in conjunction with an automatic speech
recognition system to produce a dataset for improving the robustness of natural language
understanding models to speech recognition errors.

Furthermore, synthetic data might be useful in ASR personalization \cite{yang2023text}.
The aforementioned study shows high effectives in ASR personalization using synthetic data,
in particular when there are few recordings of a speaker in the dataset.

Previous works also addressed the problem of imperfections in data produced by TTS.
Synthetic data differs from the real one in terms of naturalness and because of the
presence of artifacts. Hu et al. \cite{hu2022synt} proposed two techniques for ASR training to
alleviate the issues arising from the problems mentioned above. They observed up to
$13\%$ relative error reduction in ASR task.

The authors of VoiceBox \cite{le23voicebox} investigate the performance of ASR models
trained on real and synthetic data. For training the ASR model on real data they use
LibriSpeech 100h and 960h datasets. The synthetic data are generated from the texts
collected in the LibriSpeech training set. The evaluation is performed with respect to
\emph{test-clean} and \emph{test-other} subsets of LibriSpeech which do not contain conversational speech.
Le et al. \cite{le23voicebox} show that their best performing TTS models lead to the
absolute WER increase of $0.4\%$ on \emph{test-clean} and $1.7\%$ on \emph{test-other},
if compared to the models trained no real data.
Contrary to \cite{le23voicebox}, we investigate the impact of using voice-cloned speech on
domain-specific adaptation of ASR in the conversational setting
and use for this purpose datasets that contain conversational speech (SLURP and IVA).

\section{Data}
\label{datadescription}

Measuring the impact of synthesized data on the performance of the ASR model
requires careful selection of speech resources to be used for training and
evaluation. We decided to use LibriSpeech \cite{panayotov15librispeech} as a resource
for training baseline ASR model and as a target corpus for augmentation.
LibriSpeech is a corpus of approximately $1,000$ hours of read English speech, recorded by
more than $2,400$ speakers. It is derived from the LibriVox project, which
features audiobooks read by volunteers.

For training speech synthesizers we used LJ Speech Dataset \cite{ljspeech17} and
Hi-Fi TTS Dataset \cite{bakhturina2021hifi}. LJSpeech is a dataset of about $24$ hours
of audio from a single speaker reading book passages, specifically from
Project Gutenberg. Hi-Fi TTS Dataset is also based on Project Gutenberg texts and
LibriVox audiobooks and contains about $292$ hours of speech from $10$ speakers
with at least $17$ hours per speaker. Both of these datasets were designed for
training models for speech-based applications, with the main focus on speech synthesis.

We also utilize open-sourced VALL-E X model\footnote{\url{https://github.com/Plachtaa/VALL-E-X}}
that was trained on LibriTTS \cite{libritts}, AISHELL-1 \cite{aishell1}, AISHELL-3 \cite{aishell3}
and Japanese subset of CommonVoice dataset \cite{commonvoice}. The authors also used some
self-gathered data that was not described. In total they used about $704$ hours of speech
for English, $598$ hours for Chinese and $437$ hours for Japanese.

We evaluate ASR models using three general-purpose and two domain-specific ASR datasets.
The general-purpose datasets include two test splits of LibriSpeech, \emph{test-clean} and \emph{test-other}.
The \emph{test-clean} split has higher quality of samples compared to \emph{test-other}
\cite{panayotov15librispeech}. As a third general-purpose dataset, we use the test split of FLEURS \cite{fleurs}
which provides natural speech recordings for many languages, out of which we use an English subset only.

As for the testsets in the domain of virtual assistants, we chose to use the test split of SLURP
\cite{bastianelli20slurp} and our internal virtual assistant (IVA) dataset.
The SLURP testset has $13078$ recordings totalling $10.3$ hours of audio, while the IVA dataset
contains $14094$ recordings and $12.5$ hours of speech. IVA has a broader set of domains
and intents ($55$ and $223$ respectively) compared to SLURP ($18$ and $94$).
Table~\ref{tab:evalsets} describes the language resources used for evaluation.

\begin{table}[!ht]
\caption{Resources used for evaluation.}
\label{tab:evalsets}
\centering

\begin{tabular}{l r r r}

  \hline
  \textbf{Dataset} & \textbf{Samples} & \textbf{Hours} & \textbf{Speakers} \\
  \hline
  LS-clean & $2620$ & $5.4$ & $40$ \\
  LS-other & $2939$ & $5.1$ & $33$ \\
  FLEURS & $647$ & $1.8$ & $-$ \\
  SLURP & $13078$ & $10.3$ & $-$ \\
  IVA & $14094$ & $12.5$ & $-$ \\
  \hline

\end{tabular}

\end{table}

For prompting VALL-E X, we randomly chose one recording for each of the speakers. As sources of prompts
we used LibriSpeech, HiFi TTS Dataset and LJ Speech Dataset described above and VCTK dataset \cite{VCTK}
which contains high quality speech data recorded by $110$ English speakers.

\section{Models}

\subsection{Speech Recognition}

For our experiments we chose the Conformer on-device ASR model \cite{park23conformer}.
It is based on a RNN-Transducer architecture and has been commercialized on edge devices,
which proves its high quality. This makes it a compelling target for our experiments
on improving the ASR performance.

The model provides real time ASR performance on edge devices. Although the authors
used two pass model for better quality, we limited ourselves to the first pass.
Our main goal was to observe the difference between both augmentation approaches
so we did not find improving ASR by ensembling relevant. In our single pass approach
the transcription network encodes acoustic features of speech, while the predictor
network acts as language model and tries to predict the next token based on the previous ones.
These two, the acoustic features and language features are joined together in the
joint network that outputs the final label.

\subsection{Speech Synthesis}

As a conventional neural approach to speech synthesis we decided to use a two-stage end-to-end
TTS, consisting of an acoustic model mapping phonetic labels to acoustic features and
a vocoder mapping these features to audio samples.

The set of phonetic labels contained symbols for phonemes, word delimiters and
end of sentence marks (affirmative sentences, questions and exclamations). Acoustic
features were derived from F0 (interpolated in unvoiced regions), mel-spectra and
band-aperiodicity in a manner of the WORLD vocoder \cite{worldvocoder}. We utilized vocoder
architecture that follows LPCNet \cite{lpcnet} and an acoustic model based on Tacotron and
\cite{wang2017tacotron} Tacotron2 \cite{tacotron2}, as described in \cite{Ellinas2020High}.
For simplicity, later we refer to this system as a whole by the name LPCTron.

\subsection{Voice Cloning}

VALL-E X \cite{zhang2023vallex} is a zero-shot TTS that offers state of the art quality of
cloning a sample voice, having only a $3$-second recording of it. Instead of regarding
speech synthesis as a continuous regression task, it adopts conditional language modelling
approach, where the synthesis is conditioned on the input text and audio. It also ceases to
use mel-spectogram in favor of acoustic tokens that are generated by neural codec LM.

The output speech is modeled at two stages with a total of $8$ quantizers. In the first
stage, the autoregressive language model generates codec codes of the first quantizer. During the
second stage, the non-autoregressive language model generates codes for the rest of the quantizers,
it is conditioned not on previously generated tokens but on all the tokens from previous
quantizers. This makes the second stage much faster, because codes from previous quantizers
are known at the start. The intention is that each next quantizer encodes the details that were
not captured by previous ones.

The reason that VALL-E X is useful for our task is that it has in-context learning
ability, which means that it can synthesize high-quality output on previously unseen inputs.
While conventional neural TTS systems needed fine-tuning for unseen speakers, VALL-E X does not.

Open-source VALL-E X implementation follows the original paper \cite{wang2023neural} and uses
G2P tool for converting the input sentence to phonemes and EnCodec \cite{defossez2022high} as
a neural codec.

\section{Experiments}
\label{sec:experiments}

\begin{table*}[!ht]
\caption{LibriSpeech 960h ASR models WER.}
\label{tab:960hmodels}
\centering
\begin{tabular}{l|r|r r r r r|r r r r r}

  \hline
  \textbf{Dataset} & \textbf{BASE} & \textbf{L040} & \textbf{L060} & \textbf{L100} & \textbf{L200} & \textbf{L300} & \textbf{V040} & \textbf{V060} & \textbf{V100} & \textbf{V200} & \textbf{V300} \\
  \hline
  LS-clean & $8.08$ & $7.88$ & $7.62$ & $7.91$ & $8.07$ & $7.80$ & $7.97$ & $9.60$ & $10.74$ & $8.29$ & $8.10$ \\
  LS-other & $20.57$ & $19.84$ & $20.17$ & $20.23$ & $20.51$ & $20.58$ & $20.47$ & $21.43$ & $22.17$ & $20.79$ & $20.75$ \\
  FLEURS & $34.31$ & $34.90$ & $34.02$ & $34.04$ & $34.83$ & $34.44$ & $33.39$ & $33.72$ & $36.28$ & $35.03$ & $33.24$ \\
  SLURP & $74.89$ & $70.02$ & $69.22$ & $68.37$ & $69.56$ & $68.83$ & $66.67$ & $64.64$ & $65.56$ & $63.39$ & $62.58$ \\
  IVA & $75.14$ & $66.82$ & $64.75$ & $62.13$ & $64.09$ & $62.54$ & $50.62$ & $54.01$ & $52.91$ & $47.82$ & $44.61$ \\
  \hline

\end{tabular}

\end{table*}

The goal of our study is to investigate how does the multivoiceity of synthesized, domain-specific
training data impact the performance of the resulting ASR model. For this purpose we conduct
experiments with ASR models trained on speech recordings, speech recordings combined with data
synthesized with LPCTron and speech recordings combined with data synthesized with VALL-E X.

For synthesis, we created a text corpus consisting of $129,000$ user commands directed to a
task-oriented virtual assistant which includes $81,500$ utterances from our internal dataset,
and $47,500$ utterances obtained in the process of augmenting the training split of the SLURP dataset.

The augmentation employed to enrich SLURP consisted of two steps.
First, we used RoBERTa \cite{liu19roberta} and BART \cite{lewis20bart} models to randomly
substitute words in the user commands with their counterparts supplied by the language models.
Second, the sentences were transcribed from English to French, German,
Italian and Spanish and backwards with the use of OPUS-MT models \cite{tiedemann20opusmt}.

\begin{figure}[!ht]
\begin{center}
\includegraphics[scale=0.48]{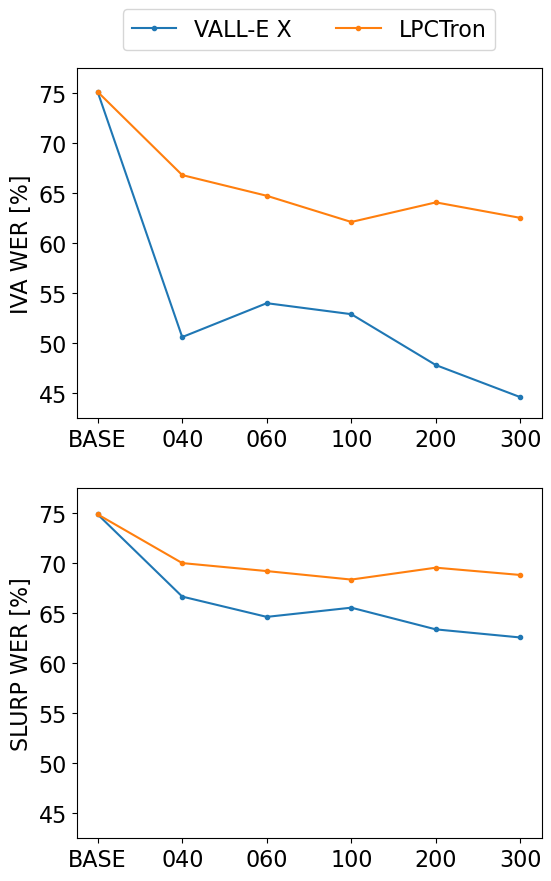}
\caption{WER obtained on SLURP and IVA.}
\label{fig:results}
\end{center}
\end{figure}

The text corpus was split into $3$ equal parts and synthesized using both LPCTron and VALL-E X. For
LPCTron we selected voices randomly from $11$ available options and for VALL-E X from $752$.
The audio prompts for VALL-E X were collected from $4$ datasets in a manner described in section
\ref{datadescription}. The first part of the text corpus was synthesized with $2$ voices per
sentence, second part with $3$ voices and the last part with $10$ voices. This way we obtained
three sets of $40$ hours, $60$ hours and $200$ hours of synthesized speech. We combined these sets
into splits: $40$ hours, $60$ hours, $100$ hours $200$ hours and $300$ hours, which were later
utilized for experiments.

We used $960$ hour subset of LibriSpeech corpus for training along with splits
of synthetic data. The \emph{Lxxx} models combine LibriSpeech recordings with LPCTron
synthesized dataset with \emph{xxx} hours, e.g. \emph{L060} used $60$ hours split
mentioned above. Analogically, the \emph{Vxxx} models combine LibriSpeech data with \emph{xxx}
hours of spoken commands generated with the use of VALL-E X model. The \emph{BASE} model is a
baseline trained on LibriSpecch $960$h without addition of synthetic data.

The results presented in Table~\ref{tab:960hmodels} indicate significant improvement in
performance of the augmented models on domain-specific testsets (SLURP and IVA).
We can also observe no significant performance drop on general-purpose test sets
(LS-clean, LS-other and FLEURS) meaning that ASR models maintained generalization
capability. The \emph{V300} performs the best out of all trained models
and results in absolute WER reduction, with regard to the \emph{BASE}, of $30.53pp$ and $12.31pp$
in comparison to $12.60pp$ and $6.06pp$ obtained by \emph{L300} on the IVA dataset and
SLURP, respectively.

To investigate how the amount of synthetic data used for training impacts the ASR,
we compared WER obtained using different data splits of IVA and SLURP. As shown in Figure~\ref{fig:results},
models trained with addition of VALL-E X data outperform
their counterparts augmented with LPCTron data.
There is also a noticeable improvement in WER with addition of more voice-cloned
data, whereas the results plateau for models trained with the usage of LPCTron data.

To verify the quality of the audio data produced by VALL-E X and LPCTron we used
Whisper \cite{radford2022robust} ASR model. We computed WER on the subset of $40$ hours
data. We got $37.55\%$ and $20.38\%$ WER on VALL-E X and LPCTron datasets, respectively.

\section{Discussion}

The choice of LPCTron as the baseline for conducting experiments can be questioned as
there are several other more recent, conventional neural TTS models that can be used
for the task. However, when comparing ratio between MOS for synthesized speech and MOS
measured for ground truth across different architectures the results for LPCTron \cite{Ellinas2020High}
$93\%$ ($=4.2/4.5$) are on par with $89\%$ ($=3.83/4.3$) achieved for FastSpeech2 \cite{ren2022fastspeech},
$98\%$ ($=4.36/4.45$) for HiFiGAN \cite{hifigan} and $93\%$ ($=3.961/4.274$) for WaveGlow \cite{Prenger_2019}.
Taking into account that HiFiGAN and WaveGlow are vocoders, not the full TTS systems,
only FastSpeech2 would be a direct replacement for LPCTron in our experimental setting.
Still, FastSpeech2 model presents similar quality to Tacotron2-based TTS models as
shown in \cite{ren2022fastspeech}. Furthermore, as we reported in Section~\ref{sec:experiments},
the transcriptions of the audio samples produced by LPCTron obtained with the use of Whisper
\cite{radford2022robust} had significantly lower WER than their VALL-E X counterparts. This
shows that the quality of generated speech was higher in the case of LPCTron making our
study sound, even if the LPCTron model is outperformed by some other conventional neural
TTS model that can be potentially used as a baseline for experiments.

Taking into consideration that the compared TTS models are trained in a different
manner with VALL-E X being trained for zero-shot (voice cloning) synthesis and
LPCTron being trained for a conventional synthesis, there are differences in the
model architecture that we cannot control in the experimental setting. However,
it should be noted that although VALL-E X is a decoder-only model and Tacotron
is an encoder-decoder model both of them are autoregressive, thus we do not consider
the differences in the architecture to have a significant impact on the results.

Before VALL-E X, other approaches to zero-shot voice-cloning speech synthesis were considered.
They were mainly based on providing the acoustic model with speaker embeddings extracted
from speech sample with speaker verification models \cite{spkverif}. This approach still relies
on the availability of high quality data for multiple speakers to train acoustic model
to utilize speaker embedding space properly. On the other hand, conditional language modelling
approach allows for utilizing lower quality data which makes it more suitable to our study.

\section{Conclusions}

In this study we investigated the efficacy of using voice-cloned speech for augmenting spoken
language with the goal of improving the performance of an ASR system. In this setting, we compared
a baseline dataset that contains solely voice recordings, the dataset with addition of
voice-cloned samples and the dataset expanded with samples synthesized by a conventional neural
TTS system.

The conducted experiments show that the use of voice cloning to generate data with multiple
voices and pronunciations improves the ASR performance significantly, compared to data from
a conventional TTS speaking in just one or a few voices. The lower quality of voice-cloned
speech, showed in terms of intelligibility, does not prevent the mentioned improvement.

We also showed that improvements gained by adding more synthetic data to the speech corpus plateau
quickly for data generated using conventional neural TTS, but adding even $300$ hours of synthetic
speech generated using VALL-E X does not seem to saturate the results of ASR model.

One avenue for further research is to investigate upper limits of augmenting speech corpora using
voice-cloned samples. Other dimension worth experimenting with is voice characteristics variability
and its impact on the ASR results. There is also noticeable gap in quality of synthesized speech in
terms of intelligibility between conventional neural TTS and LM-based TTS which should be decreased.

\bibliographystyle{IEEEtran}
\bibliography{main}

\end{document}